\documentstyle[epsfig,12pt]{article}
\newcommand{\nl}{\nonumber \\}
\newcommand{\bq}{\begin{equation}}
\newcommand{\eq}{\end{equation}}
\newcommand{\bqa}{\begin{eqnarray}}
\newcommand{\eqa}{\end{eqnarray}}
\newcommand{\eqn}[1]{Eq.(\ref{#1})}

\newcommand{\bino}[2]{\left(\begin{array}{c} #1 \\ #2 \end{array}\right)}

\begin{document}
\pagestyle{empty}
\begin{flushright}CERN-TH.6714/92\\
\end{flushright}
\vspace*{1cm}
\begin{center}\begin{Large}
{\bf Multiscalar production amplitudes}\\
{\bf beyond threshold }\\

\vspace*{2cm}

E.N.~Argyres$^*$ and
Costas~G.~Papadopoulos\\
TH Division, CERN, Geneva, Switzerland\\
\vspace{\baselineskip}
Ronald~H.P.~Kleiss,\\
NIKHEF-H, Amsterdam, the Netherlands\\
\end{Large}
\vspace*{3cm}
Abstract\end{center}
We present exact tree-order amplitudes for $H^* \to n~H$, for final
states containing one or two particles with non-zero
three-momentum, for various interaction potentials.
We show that there are potentials leading to tree amplitudes
that satisfy unitarity,
not only at threshold but also in the above kinematical
configurations and probably beyond.
As a by-product,
we also calculate $2\to n$ tree amplitudes at threshold and
show that for the unbroken $\phi^4$ theory they vanish for $n>4~$,
for the Standard Model Higgs they vanish for $n\ge 3~$ and for a model
potential, respecting tree-order unitarity, for $n$ even and $n>4~$.
Finally, we calculate
the imaginary part of the one-loop $1\to n$ amplitude in both symmetric
and spontaneously broken $\phi^4$ theory.
\vspace{1cm}
\begin{flushleft} CERN-TH.6714/92\\October 1992\\
\vfill
\begin{small}
$*$ On leave of absence from the Institute of Nuclear Physics,
NRCPS
`$\Delta  \eta  \mu  \acute{o}  \kappa  \varrho
  \iota  \tau  o  \varsigma$',
GR-153 10 Athens, Greece.
\end{small}
\end{flushleft}

\newpage
\pagestyle{plain}
\setcounter{page}{1}
\section{Introduction}

The problem of calculating amplitudes for high-multiplicity production
of weakly interacting particles has recently received considerable
attention. The tree-level contributions for $1\to n$ processes at the
threshold point (all produced particles at rest) have been calculated
exactly for $\phi^4$ scalar theory in both the symmetric case
and broken
symmetry (Standard Model Higgs) case \cite{vol1,akp1}. In addition the
one-loop contribution at threshold has also been calculated exactly
in both cases \cite{vol2,smit}, using the method proposed
in ref.\cite{brow}.

These amplitudes, growing as $n!$ with the multiplicity $n$,
lead to unitarity-violating cross sections. In a recent
paper \cite{akp3}, we discussed a potential, which allowed the
tree-order amplitudes to satisfy the unitarity bound,
at least at the treshold point.

In this paper, we extend the calculation of tree
amplitudes beyond the treshold point. We present exact results for
the cases when one or two of the produced particles have non-zero
momentum and we show that the tree amplitudes calculated using
our unitarity-respecting potential (UR potential)
continue to satisfy the
unitarity bound. This strongly suggests that the UR
potential leads to
unitarity-respecting cross sections.

As a by-product of our calculation, inspired by the work of ref.
\cite{vol2}, we also obtain the exact amplitude for two on-shell scalars
to go to $n$ scalars at rest for various potentials, with the remarkable
result that in symmetric $\phi^4$ theory all amplitudes are zero
for $n>4$ ; for the symmetry-broken case this happens for $n\ge 3$ and
for the UR potential
we get a nullification when $n$ is even and $n>4$. We also
show that this property holds for $\phi^m$ only when $m\le 4$.
Finally, using the result for the amplitude $1\to n$ when two of the
final momenta are non-zero, the above property
and the well-known Cutkosky rules, we
calculate the imaginary part of the one-loop amplitude at threshold.

The paper is organized as follows. In section 2, we discuss
the general $\phi^m$ theory,
the special case of unborken $\phi^4$ theory, and
its spontaneously broken version,
for the case where one of the final-state three-momenta is non-zero.
In section 3, we calculate tree amplitudes for
the UR potential, in the cases where one or two of the final-state
momenta are non-zero, and we show that they respect the unitarity bound.
We also derive some results for the $\phi^4$ theory.
In section 4, we calculate the imaginary part of the
one-loop amplitude for the $1\to n$ process,
in both symmetric and broken-symmetry $\phi^4$
theories. Finally, section 5 contains our conclusions.

\section{Amplitudes for final states with one non-zero momentum}

We consider the process  $H^*\to nH$
with the following configuration of the final-state momenta:
\bq
p_1^{\mu} = (E,\vec{p})\;\;\;,\;\;\;
p_i^{\mu} = (1,\vec{0})\;\;\;\mbox{for $n=2,3,\ldots,n$}\;\;,
\label{point}
\eq
where we put the mass of the Higgs equal to unity, so that
$E^2-{\vec p~}^2=1$.
Let us first consider a general potential
\bq
V(\phi) = \sum\limits_{m\ge 3}
{\lambda_{m}\over (m)!}\phi^m\;\;.
\label{potential}
\eq


The tree amplitude is given by the
following recursion relation (see fig.1):

\bqa
a(n,p_1) & = & a_1(n) \nl
& = &  -i\sum\limits_{p=2}^{n}{\lambda_{p+1}\over(p-1)!}
\sum\limits_{\begin{scriptsize}
\begin{array}{c}
n_{1},...,n_{p}\ge 1 \\  n_{1}+...+n_{p}=n
\end{array}
\end{scriptsize} }
{ia_1(n_1)\over P(n_1)}
{ia(n_2)\over (n_2^2-1)}\cdots
{ia(n_p)\over (n_p^2-1)}{n!\over n_1!n_2!\ldots n_p!},
\label{recursa}
\eqa

where $a(n)$ is
the amplitude at threshold (all outgoing momenta zero),
and $P_1(n)$ is the inverse propagator, given by
\bq
P_1(n) = \left(p_1+(n-1)p_2\right)^2-1 = (n-1)(n+\omega)\;\;;
\eq
here, we have introduced $\omega=2E-1$. The ans\"{a}tze
\bq
 a(n) = -in!(n^2-1)b(n)\;\;\;,\;\;\;
a_1(n) = -i(n-1)!P_1(n)b_1(n)\;\;,
\eq
and the introduction of the generating functions
\bq
f(x)=\sum\limits_{n\ge1}b(n)x^n\;\;\;,\;\;\;
f_1(x)=\sum\limits_{n\ge1}b_1(n)x^n\;\;  \;
\eq
transform \eqn{recursa} into the following
differental equation for $f_1(x)$:
\bq
x^2f_1''(x)+\omega xf_1'(x)-\left[\omega + V''(f(x))\right]f_1(x)
=0\;\;.
\label{diffeqn}
\eq
Note that this equation for the $f_1$ is linear; the additional,
nonlinear equation for the $f(x)$ has been discussed in
\cite{akp1,akp3}. The boundary conditions for $f_1$ are, obviously,
$f_1(0)=f(0)=0$, $f_1'(0)=f(0)=1$.
We shall now consider the solution of \eqn{diffeqn} for a number
of different potentials.

\subsection{Monomial $\phi^m$ interactions}
When $V(f)=\lambda\phi^m/m!$, we know, from \cite{akp1},
the generating function when all outgoing particles are at rest:
\bq
f(x)=x\left(1-{\lambda x^q\over2m!}\right)^{-2/q}\;\;,
\label{fzero}
\eq
where $q=m-2~$.
Thus, in this case \eqn{diffeqn} becomes
\bq
x^2f_1''(x)+\omega xf_1'(x)-\left[\omega
+{2m!\over q!}{y\over(1-y)^2}\right]f_1(x) = 0\;\;,
\eq
with $y=\lambda x^q/2m!$. We now introduce $G(y)$ and $r>0$
as follows:
\bq
f_1(x)=x(1-y)^rG(y)\;\;\;,\;\;\;r(r-1)={2m!\over q^2q!}\;\;,
\label{reqn}
\eq
to find
\bq
y(1-y)G''(y) + \left\{{q+1+\omega\over q}
-\left[2r+{1+\omega\over q}+1\right]y\right\}G'(y)
-r\left(r+{1+\omega\over q}\right)G(y) = 0\;\;,
\eq
with boundary condition $G(0)=1$. This is the hypergeometric
equation \cite{abra}. The resulting solution for $f_1$ can be
written as
\bqa
f_1(x)& = & x\left(1-{\lambda x^q\over2m!}\right)^{1-r}
F\left(1-r,{q(1-r)+\omega+1\over q};{q+\omega+1\over q};
{\lambda x^q\over2m!}\right)\;\;,\nl
F(a,b;c;t) & \equiv & \sum\limits_{n\ge0}
{(a)_n(b)_n\over(c)_n}{x^n\over n!}\;\;,
\eqa
where $(a)_n=(a+n-1)!/(a-1)!$ is the Pochhammer symbol.
If it happens that $r-1$ is an integer, the summation will
end at $n=r-1$ and the hypergeometric function $F$ is a polynomial
of degree $r-1$. The denominator $((\omega+1)/q+1)_n$ vanishes
whenever $\omega=-qk-1$ for $k=1,2,\ldots,r-1$.

Having determined $f_1(x)$ we can, in principle, get the $b_1(n)$
by expanding in powers of $x$, and this gives us the amplitudes
$a_1(n)$. This $1\to n$ amplitude is equal, by crossing, to
the $2\to n-1$ amplitude, if we take for $p_1^{\mu}$ an unphysical
value with negative energy. Since $n$ must, in a $\phi^m$ theory,
be of the form $qk+1$, this corresponds to taking $\omega=-qk-1$.
Then, $a_1(qk+1)$ will contain the factor $P_1(qk+1)$ and will hence be
zero - {\em unless\/} this zero is cancelled by a corresponding simple
pole in the function $f_1(x)$. It follows that the only $a_1(qk+1)$
that are non-zero (with the above choice for $\omega$) are those with
$k\le r-1$, provided, of course, that $r$ is indeed an integer.
Since \eqn{reqn} implies $r=2(m-1)/(m-2)$, we see that this can only be
the case if $m=3$ $(r=4)$ or $m=4$ $(r=3)$. We conclude that in
a pure $\phi^3$ theory all threshold amplitudes $2\to n$ with
$n>3$, and in a pure $\phi^4$ theory all such amplitudes with $n>4$,
will vanish. In theories with higher purely monomial interaction,
no such `nullification' takes place.\\

In the special case of a $\phi^4$ theory, we find the following
explicit results:
\bq
f_1(x)  =  \left(1-{\lambda_4x^2\over48}\right)^{-2}
\left\{x+{2(3-\omega)\over3+\omega}\left(
{\lambda_4\over48}\right)x^3
+{(3-\omega)(1-\omega)\over(3+\omega)(5+\omega)}
\left({\lambda_4\over48}\right)^2x^5\right\}\;\;,
\label{fonedef}
\eq
and
\bqa
a_1(3) & = & -i\lambda_4\;\;,\nl
a_1(2k+1) & = & -i(2k)!(2k)(2k+1+\omega)
\left({\lambda_4\over48}\right)^k  \nl & & \times
\left\{k+1+{2(3-\omega)\over(3+\omega)}k
+{(3-\omega)(1-\omega)\over(3+\omega)(5+\omega)}(k-1)\right\}\;\;,
\eqa
where $k\ge2$ in the last line. It is easily checked that
$a_1(n)$ reduces to $a(n)$ when $\omega\to1$, as it should.
Moreover, by letting $\omega$ approach $-2k-1$ we
immediately find the following $2\to n$ amplitudes:
\bqa
a(2\to2) & = & -i\lambda_4\;\;,\nl
a(2\to4) & = & i\lambda_4^2\;\;,\nl
a(2\to n) & = & 0\;\;\;\mbox{for all $n>4$}\;\;.
\label{phifour}
\eqa
These are the threshold zeros noted in ref.\cite{vol2}. Note
that we have obtained, in addition to those results, the explicit
form of the amplitudes.

\subsection{The Minimal Standard Model}
In the Minimal Standard Model, with spontaneous symmetry breaking,
we have the potential
\bq
V(\phi) = \sqrt{{\lambda_4\over12}}\phi^3
+ {\lambda_4\over24}\phi^4\;\;,
\eq
and, from \cite{akp1}:
\bq
f(x) = x\left(1-x\sqrt{{\lambda_4\over12}}\right)^{-1}\;\;,
\eq
so that \eqn{diffeqn} becomes
\bq
x^2f_1''(x) + \omega xf_1'(x)
-\left[\omega+{6y\over1-y}+{6y^2\over(1-y)^2}\right]f_1(x) = 0 \;\;,
\eq
with $y=x\sqrt{\lambda_4/12}$. We now put $f_1(x)=x(1-y)^3G(y)$,
to find the equation
\bq
y(1-y)G''(y)+\left(2+\omega
-[(4+\omega)+4]y\right)G'(y)-3(4+\omega)G(y) = 0\;\;,
\eq
which is again of the hypergeometric type, this time with $b=3$.
We find, for the generating function:
\bq
f_1(x) = x\left(1-x\sqrt{{\lambda_4\over12}}\right)^{-2}
\left\{1+\sqrt{{\lambda_4\over3}}{1-\omega\over2+\omega}x
-{\lambda_4\over12}{\omega(1-\omega)\over(2+\omega)(3+\omega)}x^2
\right\}\;\;,
\label{fone}
\eq
and for the tree amplitude:
\bqa
a_1(n) & = & -i(n-1)!(n-1)(n+\omega)
\left({\lambda_4\over12}\right)^{(n-1)/2}\nl & & \times
\left\{n+2(n-1){1-\omega\over2+\omega}
-{\omega(1-\omega)\over(2+\omega)(3+\omega)}\right\}\;\;.
\eqa
As a check of this result, observe that indeed
\bq
\lim_{\omega\to1}a_1(n) = -in!(n^2-1)\left(
{\lambda_4\over12}\right)^{(n-1)/2}
=a(n)\;\;,
\eq
as it should. Letting again $\omega\to-n$, we get for the threshold
amplitudes
\bqa
a(2\to1) & = & -i\sqrt{3\lambda_4}\;\;,\nl
a(2\to2) & = & 4i\lambda_4\;\;,\nl
a(2\to n) & = & 0\;\;\;\mbox{for all $n\ge3$}\;\;.
\eqa
These are the threshold zeros of ref.\cite{smit}.

\subsection{Unitarity-respecting model potential}
In a previous publication \cite{akp3}, we studied a model potential
that allows the tree-level amplitudes to satisfy unitarity at
the threshold point. We now show that unitarity is also respected
at phase-space points of the type of \eqn{point}. The UR
model has a
potential $V(\phi)$, and a generating function $f(x)$,
given by
\bqa
V(\phi) & = & {1\over2}(1+\phi)^2\left(\log(1+\phi)\right)^2\;\;,\nl
f(x) & = & e^x-1\;\;,
\eqa
and, consequently,
\bq
V''(f) = x^2+3x\;\;.
\eq
Inserting this into \eqn{diffeqn}, and writing
\bq
f_1(x) = xe^{-x}G(y)\;\;\;,\;\;\;y=2x\;\;,
\eq
we get the following equation for $G$:
\bq
yG''(y) + (\omega+2-y)G'(y)+{\omega+5\over2}G(y)=0\;\;,
\eq
whose solution is the confluent hypergeometric function \cite{abra}.
The final solution with boundary conditions $f_1(0)=0$ and
$f_1'(0)=1$ is
\bq
f_1(x) = xe^{-x}M\left({\omega+5\over2},\omega+2;2x\right)\;\;,
\label{modelfun}
\eq
where $M$ is a Kummer function \cite{abra}:
\bq
M(a,b;z) = \sum\limits_{n\ge0}{(a)_n\over(b)_n}{z^n\over n!}\;\;.
\label{kummer}
\eq
Since this series has an infinite radius of convergence,
so does $f_1(x)$, and hence the coefficients $b_1(n)$ decrease
sufficiently fast with $n$ to satisfy unitarity.
In fact, we can read off the explicit form for $b_1(n)$ (and hence
that for $a_1(n)$ immediately from \eqn{modelfun}: after some trivial
algebra, we obtain
\bq
a_1(n)= -i(n-1)(n+\omega)\sum\limits_{k=0}^{n-1}
(-1)^{n-1-k}\bino{n-1}{k}\prod\limits_{j=2}^{k+1}
{\omega+2j+1\over\omega+j}\;\;.
\eq
Note that we define the empty product for $k=0$ to be unity.
The check that $a_1(n)\to n^2-1$ as $\omega\to 1$ follows immediately,
and we also find, by letting $\omega\to-p-1$ with integer $p$,
that $a_1(p+1)$, and therefore the threshold amplitude for
the $2\to p$ process, will vanish whenever $p$ is even and larger
than 4. So also in this case, there is  nullification. Note, however,
that the technical reason for this nullification is slightly different
from that in the polynomial scalar theories: there, the fact that the
hypergeometric function turns out to be a finite polynomial limits the
number of different poles, whereas in this case some of the poles cancel
against corresponding numerators, without the Kummer function
having to reduce to a finite polynomial (although, in fact, it does).

Before finishing this section, we want to remark that we have also
reproduced the threshold nullification described in ref.\cite{vol3},
for the annihilation of a fermion-antifermion pair, and of
a vector boson pair, into scalars. In addition, we have found explicit
forms for the amplitudes. Since these processes, however, involve
more than just scalar particles, we defer their study to a forthcoming
publication \cite{akpnew}.

\section{Final states with two non-zero momenta}
\subsection{Back-to-back momenta}
We now turn to the case where two of the final-state momenta have
nonvanishing space-like components. As a first step, we take the
momentum configuration
\bq
p_1^{\mu}=(E,\vec{p})\;\;,\;\;
p_2^{\mu}=(E,-\vec{p})\;\;,\;\;
p_i^{\mu}=(1,\vec{0})\;\;,\;\;(i=3,4,\ldots,n)\;\;.
\label{twospecial}
\eq


For a generic potential of the form (\ref{potential}),
the tree-level amplitude $a_2(n)$ now obeys (as depicted in
fig.2) the following inhomogeneous recursion relation:
\bqa
ia_2(n) & = &
\sum\limits_{p=2}^n{\lambda_{p+1}\over(p-1)!}
\sum\limits_{n_1,\ldots,n_p \ge 1}
{ia_2(n_1)\over P_2(n_1)}\left\{
{ia(n_2)\over n_2^2-1} \cdots \right.  \nl
& & \hspace{2cm}\left.\cdots
{ia(n_p)\over n_p^2-1}
{(n-2)!\over(n_1-2)!n_2!\cdots n_p!}
\delta_{n_1+\cdots+n_p,n}\right\}\nl
       & + &
\sum\limits_{p=2}^n{\lambda_{p+1}\over(p-2)!}
\sum\limits_{n_1,\ldots,n_p \ge 1}
{ia_1(n_1)\over P_1(n_1)}
{ia_1(n_2)\over P_1(n_2)}\left\{
{ia(n_3)\over n_3^2-1}\cdots    \right.\nl
 & & \hspace{2cm}\left.\cdots
{ia(n_p)\over n_p^2-1}
{(n-2)!\over(n_1-1)!(n_2-1)!n_3!\cdots n_p!}
\delta_{n_1+\cdots+n_p,n}\right\}\;\;,
\label{recursatwo}
\eqa
where
\bqa
P_2(n) & = & (p_1+p_2+(n-2)p_3)^2-1  \nl
& = & (n+\rho-1)(n+\rho+1)\;\;\;,\;\;\;\rho=2(E-1)\;\;.
\label{propa}
\eqa
Our ansatz for $a_2(n)$ is
\bq
a_2(n) = -i(n-2)!P_2(n)b_2(n)\;\;;
\eq
with this, and the following definition of the generating function:
\bq
f_2(x) = \sum\limits_{n\ge2}b_2(n)x^{n+\rho}\;\;,
\label{genfun}
\eq
we find the analogous form of \eqn{diffeqn}:
\bq
x^2f_2''(x) + xf_2'(x) - \left[1+V''(f(x))\right]f_2(x)
-x^{\rho}f_1(x)^2V'''(f(x)) = 0\;\;.
\label{difftwo}
\eq
In our model UR potential, we have
\bq
V'''(f(x))=e^{-x}(2x+3)\;\;,
\eq
so, upon substituting the results of the previous section, we
have the following inhomogeneous differential equation:
\bq
x^2f_2''(x) + xf_2'(x) - (1+3x+x^2)f_2(x) = F(x)\;\;,
\eq
with
\bq
F(x)=x^{2+\rho}e^{-3x}(2x+3)M(2+E,2E+1,2x)^2\;\;.
\eq
The boundary conditions that we have to take are
\bq
\lim\limits_{x\to0} x^{-\rho}f_2(x) =
\lim\limits_{x\to0}{d\over dx}\left( x^{-\rho}f_2(x)\right)
 = 0\;\;.
\eq
The corresponding solution is given by
\bq
f_2(x) = -4xe^{-x}\int\limits_{0}^{x}\;dt\;
\left\{e^{2t}U(3,3;2x)-e^{2x}U(3,3;2t)\right\}e^{-t}F(t)\;\;,
\label{difficult}
\eq
where $U$ is the {\em singular\/} solution of the confluent
hypergeometric equation \cite{abra}:
\bq
U(3,3;2x) = {1\over2}e^{2x}E_1(2x)-{1\over4x}+{1\over8x^2}\;\;,
\eq
and $E_1$ is the exponential integral
\bq
E_1(z) = \int\limits_{z}^{\infty}dt\;{e^{-t}\over t}
= - \gamma - \log(2x) +
\sum\limits_{n\ge1}{(-2x)^n\over n!n}\;\;.
\eq
As discussed in \cite{akp3}, if $f_2(x)$, when expanded
in powers of $x$, has an infinite radius of convergence, the amplitude
$a_2(n)$ will not grow factorially with $n$, and hence presumably
satisfy unitarity. From \eqn{difficult} this is, however,
not evident, since the function $U$ has a logarithmic
singularity as well as a pole at $x=0$. To see that
these singularities actually do
cancel, consider first the contribution to $f_2(x)$
coming from the logarithmically singular part (LS):
\bqa
f_2(x)_{LS} & = & \int\limits_{0}^{x}\;dt\;
\left\{e^{2t}\left(-{1\over2}e^{2x}\log x\right)
-e^{2x}\left(-{1\over2}e^{2t}\log t\right)\right\}
e^{-t}F(t)\nl
& = & -{1\over2}e^{2x}\sum\limits_{n\ge0}
{u_n\over(n+1+\rho)}x^{n+1+\rho}\;\;,
\eqa
where we wrote $e^{-t}F(t)=\sum_{n=2}^{\infty}u_nt^n$, and
integrated by parts. Thus, the logarithmically singular part
gives a regular contribution, and the result for $f_2(x)$ has
an infinite radius of convergence. As concerns the pole terms,
it is easy to see that they do not spoil the convergence of the
integral since the function $F(t)$ has leading behaviour $t^{2+\rho}$.
We have therefore shown that the amplitudes do not exhibit
unitarity-violating growth, also at the phase-space points
defined by \eqn{twospecial}.\\

Another case, which will be of interest later on in the calculation
of the one-loop correction,
is that of the pure $\phi^4$ theory, with $p_1^0=p_2^0=2$;
the lowest non-zero $b_2(n)$ is $b_2(3)$.
In this case we have $\rho=2$, and in the $f_1(x)$
given in \eqn{fone} we have, of course, to use $\omega=3$.
Inserting the results (\ref{fzero}) and (\ref{fone})
into \eqn{difftwo} we find
\bq
x^2f_2''(x) + xf_2'(x)-f_2(x) =
{\lambda_4\over2}{x^2\over(1-\lambda_4x^2/48)^2}f_2(x)
+\lambda_4{x^5\over(1-\lambda_4x^2/48)^5}\;\;.
\eq
The (simple) solution to this that starts with $x^5$ is
\bq
\phi(x) = {\lambda_4\over24}{x^5\over(1-\lambda_4x^2/48)^3}\;\;.
\eq
(Note that the same solution, albeit with different normalization,
also occurs in ref.\cite{vol2}). The explicit form of $b_2$
follows immediately:
\bq
b_2(2k+1) = k(k+1)\left({\lambda_4\over48}\right)^k\;\;.
\label{btwo}
\eq

\subsection{General momenta}
The above considerations can be generalized to evaluate the tree
amplitude when two of the outgoing particles have more general momenta:
\bq
p_1^{\mu}=(E_1,\vec{p}_1)\;\;\;,
p_2^{\mu}=(E_2,\vec{p}_2)\;\;\;,
p_i^{\mu}=(1,\vec{0})\;\;(i=3,\ldots,n)\;\;.
\eq
Let us describe this case briefly. The inverse propagator,
$P_2(n)$, is still defined as in \eqn{propa}, but now has the form
\bqa
P_2(n) & = & (n+\rho-1)(n+\rho+\alpha)\;\;,\nl
\rho & = & E_1+E_2-1-
\left(E_1^2+E_2^2+
2\vec{p}_1\!\cdot\!\vec{p}_2-1\right)^{1/2}\;\;,\nl
\alpha & = &-1 + 2
\left(E_1^2+E_2^2+
2\vec{p}_1\!\cdot\!\vec{p}_2-1\right)^{1/2}\;\;.
\eqa
The differential equation for the generating function (\ref{genfun})
now takes the form
\bq
x^2f_2''(x) + \alpha xf_2'(x) -
(\alpha+x^2+3x)f_2(x) = G(x)\;\;,
\eq
where the inhomogeneous term is
\bq
G(x) = x^{2+\rho}e^{-3x}(2x+3)
M(E_1+2,2E_1+1;2x)M(E_2+2,2E_2+1;2x)\;\;,
\eq
and the boundary conditions are as before. The solution is
\bq
f_2(x) = Axe^{-x}\int\limits_{0}^{x}
\;dt\;t^{\alpha-1}e^{-t}\left\{
\Phi_1(t)\Phi_2(x)-\Phi_1(x)\Phi_2(t)\right\}G(t)\;\;,
\eq
with
\bqa
\Phi_1(x) & = & M\left({5+\alpha\over2};2+\alpha;2x\right)\;\;,\nl
\Phi_2(x) & = & U\left({5+\alpha\over2};2+\alpha;2x\right)\;\;,\nl
A & = & -2^{\alpha+1}{\Gamma\left({5+\alpha\over2}\right)\over
\Gamma(2+\alpha)}\;\;.
\eqa
Noting that $t^{\alpha-1}G(t)\sim t^{\alpha+1+\rho}$,
and that the singular part of $\Phi_2(x)$ goes as $t^{-1-\alpha}$,
we can easily show that the solution $f_2(x)$ has again a series
expansion with an inifinite radius of convergence, with the usual
implication for the high-$n$ behaviour of $a_2(n)$ for this
phase-space point.

\section{On the one-loop correction at threshold}
A by-product of the above calculation is that it enables
us to obtain, without much effort, the imaginary part of the
one-loop amplitude $a(n)$. Since, as we showed in
\eqn{phifour}, for the $\phi^4$ potential the only non-zero
$2\to n$ threshold amplitudes are those for $n=2$ and $n=4$, and
since the $2\to2$ process has vanishing phase space, application
of the Cutkosky rule (see fig.3) gives


\bq
\mbox{Im}\;a(n)_{\mbox{{\small 1-loop}}} =
{1\over2}\left[a_2(n-2;p_1,p_2)\right]_A
\bino{n}{4}V_2(p_1,p_2)
\left[a_{2\to4}(p_1,p_2)\right]_{OS}\;\;.
\label{cutcos}
\eq
Here, $a_2(n-2;p_1,p_2)_A$ is the amputated tree amplitude for
the $1\to n-2$ transition in the case studied in section
3.1, where $p_1$ and $p_2$ are back-to-back, and the other
$n-4$ momenta at rest. $V_2$ is the two-particle phase space volume.
The amplitude $a_{2\to4}(p_1,p_2)_{OS}$ describes the
transition of two on-shell particles, with momenta $p_1$ and $p_2$,
to four particles at rest; and the binomial factor counts the
different numbers of ways in which four particles can be selected to
be attached to the loop. The factor $1/2$ comes from the optical theorem,
which relates the imaginary part to half of the discontinuity.

Because of energy-momentum conservation in the $2\to4$ part of the
amplitude (recall that we only consider the cut part of the diagram, in
which the intermediate states have to be put on shell), we have,
in the rest frame of $p_1+p_2$, $p_1^0=p_2^0=2$. Thus, the phase-space
factor amounts to
\bq
V_2(p_1,p_2) = {1\over2(2\pi)^2}
\int{d^3\vec{p}_1\over2p_1^0}
{d^3\vec{p}_2\over2p_2^0}
\delta^3(\vec{p}_1+\vec{p}_2)\delta(p_1^0+p_2^0-4)
= {\sqrt{3}\over32\pi}\;\;,
\eq
where we have to include a factor $1/2$ for the Bose
symmetry of the 2-boson final state.
The $2\to4$ amplitude has already been given in \eqn{phifour}.
Moreover, using the result of section 3.1, we have
\bq
b_2(n-2) = {1\over4}(n-1)(n-3)
\left({\lambda_4\over48}\right)^{(n-3)/2}
= k(k-1)
\left({\lambda_4\over48}\right)^{k-1}
\;\;\;,\;\;\;n=2k+1\;\;.
\eq
Hence, we find for the last remaining ingredient, the amputated
amplitude:
\bq
a_2(n-2;p_1,p_2)_A = {a_2(n-2)\over P_2(n-2)}
= -i(2k-3)!k(k-1)\left({\lambda_4\over48}\right)^k\;\;.
\eq
Putting everything together, and inserting the correct power of
the mass $m$,
we find for the imaginary part
of the one-loop corrected amplitude:
\bq
\mbox{Im}\;a(n)_{\mbox{{\small 1-loop}}} =
(2k+1)!\left({\lambda_4\over48m^2}\right)^k
k(k-1)\lambda_4{\sqrt{3}\over32\pi}\;\;.
\eq
Since for the spontaneously broken theory the only
non-zero amplitudes ($2\to1$ and $2\to2$) have vanishing phase
space, in that case the imaginary part is of course zero.
We find ourselves in exact agreement with refs.\cite{vol2,smit}
(note that our $\lambda_4$ differs from the convention used there
by a factor of 6).

\section{Conclusions}
We have shown that the calculation of the tree amplitudes in scalar
theories can be extended beyond the threshold point, namely to cases
where one or two of the produced particles have non-zero three-momentum.
Applied to our unitarity-respecting model potential, the
amplitudes satify unitarity also at these, slightly more
general, phase space points.

We have confirmed the results of Voloshin\cite{vol2} and
Smith\cite{smit} that
certain $2\to n$ threshold amplitudes become zero. Our treatment
is more general than theirs,
in that we have derived explicit expressions
for the amplitudes, also in cases where no `nullification'
occurs. We have shown that the nullification for purely monomial
interactions is restricted to the $\phi^3$ and $\phi^4$ cases.
In addition, we have shown that in our toy potential a similar
nullification occurs for $n$ even and larger than 4.

Finally, we have applied
our explicit results for the various amplitudes,
and a simlpe Cutkosky rule, to rederive the imaginary part of
the one-loop correction to the threshold point studied by
Voloshin\cite{vol2} and Smith\cite{smit}.
Although with our technique we can only reproduce
the imaginary, finite, part of the correction, we find that
the derivation presented above is quite direct and gives a better
idea of the physics behind this result.

\section*{Acknowledgements}
E. N. Argyres and C. G. Papadopoulos
are partially supported by the EEC Program SC1-C   T91-0729.\\

\end{document}